\documentclass[letter,12pt]{article}
\usepackage[top=1in,bottom=1in,left=1in,right=1in]{geometry}
\usepackage[T1]{fontenc} 
\usepackage{slashed}
\usepackage{epsfig}
\usepackage{comment}
\usepackage{cancel}
\usepackage{bbm}
\usepackage{array}
\usepackage{bigints}
\usepackage{booktabs}
\usepackage{color}
\usepackage{dsfont}
\usepackage{float}
\usepackage{framed}
\usepackage{ulem}
\usepackage{graphicx}
\usepackage{indentfirst}
\usepackage{mathrsfs}
\usepackage{multirow}
\usepackage[mathscr]{eucal}
\usepackage{subdepth}
\usepackage{titlesec}
\usepackage[dotinlabels]{titletoc}
\usepackage{wrapfig}
\usepackage[all]{xy}
\usepackage[vcentermath]{youngtab}
\usepackage{relsize}
\usepackage{hyperref}
\usepackage{empheq}
\usepackage{tikz}
\usepackage{tikz-feynman,contour}
\usetikzlibrary {positioning}
\definecolor {processblue}{cmyk}{0.96,0,0,0}
\usetikzlibrary{arrows.meta}

\numberwithin{equation}{section}

\usepackage[utf8]{inputenc}
\usepackage{slashed}
\usepackage{amsmath}
\usepackage{amsfonts}
\usepackage{amsthm}
\usepackage{amssymb}
\usepackage{cite}
\usepackage{mathtools}
\usepackage[utf8]{inputenc}

\usepackage{cleveref}
\crefname{section}{§}{§§}
\Crefname{section}{§}{§§}

 \def\p{\partial}
 \def\bz{{\bar z}}
 \def\bw{{\bar w}}
 
\def\0{{(0)}}
\def\1{{(1)}}
\def\2{{(2)}}

 \def\CL{{\cal L}}

\def\ci{{\mathscr I}}

\def\<{\langle }
\def\>{\rangle }
\def\bw{{\bar w}}
\newcommand{\bea}{\begin{eqnarray}}
\newcommand{\eea}{\end{eqnarray}}
\newcommand{\be}{\begin{equation}}
\newcommand{\ee}{\end{equation}}
\newcommand{\ba}{\begin{align}}
\newcommand{\ea}{\end{align}}

   \makeatletter
  \let\over=\@@over \let\overwithdelims=\@@overwithdelims
  \let\atop=\@@atop \let\atopwithdelims=\@@atopwithdelims
  \let\above=\@@above \let\abovewithdelims=\@@abovewithdelims
\renewcommand\section{\@startsection {section}{1}{\z@}%
                                   {-3.5ex \@plus -1ex \@minus -.2ex}
                                   {2.3ex \@plus.2ex}%
                                   {\normalfont\large\bfseries}}

\renewcommand\subsection{\@startsection{subsection}{2}{\z@}%
                                     {-3.25ex\@plus -1ex \@minus -.2ex}%
                                     {1.5ex \@plus .2ex}%
                                     {\normalfont\bfseries}}

\newcommand{\beq}{\begin{equation}}
\newcommand{\eeq}{\end{equation}}
\newcommand{\beqa}{\begin{eqnarray}}
\newcommand{\eeqa}{\end{eqnarray}}
\newcommand{\beqar}{\begin{eqnarray*}}

\def\[{\big[}
\def\]{\big]}

\def\ra{\rangle}

\def\G{\Gamma}

\def\g{{\gamma}}

\def\Th{{\Theta}}

\def\w{\omega}

\def\bz{{\bar z}}

\def\be{{\bar \epsilon}}

\def\bw{{\bar w}}

\def\CD{{\mathcal D}}

\def\CL{{\mathcal L}}

\def\CN{{\mathcal N}}
\def\CO{{\mathcal O}}

\def\+{{(+)}}
\def\-{{(-)}}
\def\0{{(0)}}
\def\1{{(1)}}
\def\2{{(2)}}
\def\3{{(3)}}
\def\4{{(4)}}
\def\5{{(5)}}

\definecolor{vecolor}{rgb}{0.7,0.3,0.9}

\def\cc{\text{c.c.}}

\def\ci{{\mathscr{I}}}

\def\amp{{\alpha}}
\def\vac{{\text{vac}}}

\begin{document}
\begin{titlepage}
\unitlength = 1mm
\hfill CALT-TH 2023-013

\ \\
\vskip 3cm
\begin{center}

{\LARGE{From Shockwaves to the Gravitational Memory Effect}}

\vspace{0.8cm}
Temple He$^1$, Ana-Maria Raclariu$^2$, Kathryn M. Zurek$^1$

\vspace{1cm}

{\it  $^1$Walter Burke Institute for Theoretical Physics \\ California Institute of Technology, Pasadena, CA 91125 USA}\\
{\it  $^2$Institute for Theoretical Physics, University of Amsterdam,
Science Park 904, Postbus 94485, 1090 GL Amsterdam, The Netherlands}

\vspace{0.8cm}

\begin{abstract}
We study the relationship between shockwave geometries and the gravitational memory effect in four-dimensional asymptotically flat spacetime.  In particular, we show the 't Hooft commutation relations of shockwave operators are equivalent to the commutation relation between soft and Goldstone modes parametrizing a sector of the gravitational phase space.  We demonstrate this equivalence via a diffeomorphism that takes a shockwave metric to a metric whose transverse traceless component is the gravitational memory. The shockwave momentum in 't Hooft's analysis is related to the soft graviton mode, which is responsible for the memory effect, while the shift in the shockwave position is related to the Goldstone mode. This equivalence opens new directions to utilize the gravitational memory effect to explore the observational implications of shockwave geometries in flat space.
 \end{abstract}

\vspace{1.0cm}
\end{center}
\end{titlepage}
\pagestyle{empty}
\pagestyle{plain}
\pagenumbering{arabic}

\tableofcontents

\section{Introduction}

The gravitational memory effect measures the net relative displacement between asymptotic detectors induced by the passage of gravitational waves  \cite{zeldovich,Braginsky:1985vlg,thorne, Ludvigsen:1989kg, Christodoulou:1991cr, Wiseman:1991ss, PhysRevD.45.520}. More recently, in the context of high energy physics, it was shown that the memory effect can equivalently be understood as the leading soft graviton theorem in particle scattering amplitudes, as well as a vacuum transition between two inequivalent vacua in asymptotically flat spacetimes \cite{Strominger:2014pwa} (see \cite{Strominger:2017zoo} for a review).

In a complementary series of developments, 't Hooft proposed that incoming and outgoing modes near black hole horizons  obey a set of commutation relations \cite{tHooft:1996rdg} and used these to evaluate a universal contribution to the black hole S-matrix. This takes the simple form of a phase shift and is directly related to the time delay acquired by a particle crossing a shockwave. The same contribution was later shown to be reproduced by resumming an infinity of graviton exchanges in the eikonal/high-energy limit of any 2-to-2 gravitational scattering process \cite{tHooft:1987vrq, Amati:1987uf, Verlinde:1991iu,Kabat:1992tb}. These results triggered a fruitful, ongoing program of understanding black holes from shockwave dynamics \cite{Kiem:1995iy, Shenker:2013pqa, Akhoury:2013bia, Shenker:2013yza, Cornalba:2006xk,Polchinski:2015cea,Ahn:2019rnq,Gaddam:2020mwe,Gaddam:2021zka,Gaddam:2022pnb}. More recently, the 't Hooft commutation relations were utilized in a proposal for how quantum gravitational fluctuations could be observed in table-top interferometers \cite{Verlinde:2019xfb,Verlinde:2022hhs,Zhang:2023mkf}.

A priori, the gravitational memory effect and the time delay experienced by a particle propagating in a shockwave background appear to be very different physical phenomena. Indeed, through its relation with the soft theorems, the gravitational memory qualifies as an infrared effect, while the time delay is revealed in the high-energy limit of scattering processes and therefore more naturally associated with the ultraviolet. The goal of this paper is to provide evidence that the two effects are nevertheless closely related. In particular, we find a diffeomorphism that relates a family of shockwave metrics to a family of Bondi metrics exhibiting a leading order memory effect. Restricting to the linearized theory is sufficient to establish this relation. This allows us to identify the shockwave variables appearing in the 't Hooft commutation relations with the soft and Goldstone\footnote{In this paper, we will interchangeably refer to these jointly as soft variables, modes, or operators.} variables \cite{He:2014laa} parametrizing the gravitational phase space of asymptotically flat spacetimes. We show that the 't Hooft commutation relations then follow directly from the commutation relations of the soft modes.

More specifically we consider a shockwave background in four spacetime dimensions.  Such shocks have been previously studied in a variety of contexts \cite{Aichelburg,dray-hooft,Sfetsos:1994xa, tHooft:1996rdg, Polchinski:2015cea,Cristofoli:2020hnk}, and are solutions to the Einstein field equations with a delta function source. We will focus on the  radial shock sourced by a radially expanding shell of massless particles.\footnote{As indicated in many places in the literature, {\it e.g.} \cite{dray-hooft}, the spherical shock cannot be sourced by a single particle.}  The spherical shock reduces to the more familiar planar shock for two points with small transverse separation on the celestial sphere.

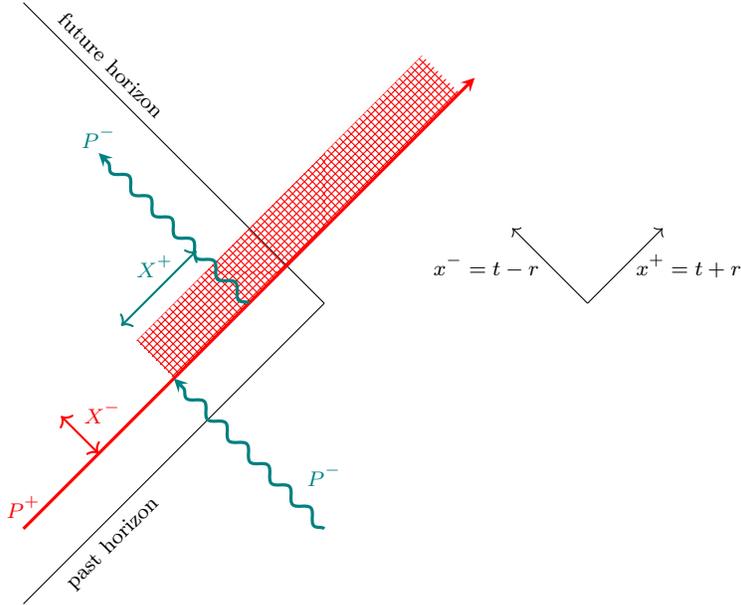
\begin{figure}[t]\label{fig:shockwave}
\centering
\begin{tikzpicture}
[decoration={markings, mark= at position 0.5 with {\arrow{stealth}}}] 
   \draw[-] (0,0) -- (4,4) node[below, near start, rotate=45] {\scriptsize past horizon};
   \draw[-] (4,4) -- (0,8) node[above, near end, rotate=-45] {\scriptsize future horizon};
   \draw[->] (7.5,4) -- (6.5,5);
   \node[] at (7, 4.5) [left] {\scriptsize $x^- = t- r $};
   \draw[->] (7.5,4) -- (8.5,5);
   \node[] at (8,4.5) [right] {\scriptsize $x^+ = t+ r $};
   \draw[-stealth,very thick, red] (0,1) -- (6,7);
   \draw[<->, thick, red] (1,2) -- (0.5,2.5) node[midway,above] {\hspace{0.6cm}\scriptsize $X^-$};
   \node[red] at (0,1) [above] {\scriptsize $P^+$};
   \draw [->,very thick, teal, decorate,decoration={snake,amplitude=0.7mm,segment length=4mm}, -stealth] (4,1) -- (2,3);
   \fill[pattern = crosshatch, pattern color=red] (2,3) -- (1.5,3.5) -- (5.3,7.3) -- (5.8,6.8) -- cycle;
   \node[teal] at (4,2) [below] {\scriptsize $P^-$};
   \draw [->,very thick, teal, decorate,decoration={snake,amplitude=0.7mm,segment length=4mm}, -stealth] (3,4) -- (1,6);
   \node[teal] at (1,6.5) [below] {\scriptsize $P^-$};
   \draw[<->, thick, teal] (1.3,3.7) -- (2.3,4.7) node[midway,above] {\hspace{-0.1cm}\scriptsize $X^+$};
   \end{tikzpicture}
\caption{\small We depict the spherical shockwave spacetime in lightcone coordinates $x^- = t-r = u$ and $x^+ = t+r = u+2r$. The particle (or rather, shell of particles) generating the shockwave, depicted in red, is localized at $x^- = x_0^-$ and $z=z_0$ (the transverse directions are suppressed). The probe particle crossing the shockwave experiences a null shift $X^+$ that is related to the shockwave momentum $P^+$. The shockwave also experiences a $X^-$ shift due to the momentum $P^-$ of the probe.}
\end{figure}

In the context of Schwarzschild horizons, 't Hooft argued that quantum mechanically there is an inherent uncertainty between the locations of ingoing and outgoing modes of Hawking particles \cite{tHooft:1990fkf, tHooft:1996rdg}. Here the ingoing probe particle experiences a shift $X^+$ due to the shockwave generated by an outgoing particle with momentum $P^+$, while the outgoing shockwave similarly experiences a shift $X^-$ due to the ingoing probe particle's momentum $P^-$. This uncertainty is expressed by a nontrivial commutation relation between ingoing and outgoing modes, as shown in Figure~\ref{fig:shockwave}, which takes the form
\begin{equation}
\label{eq:hooftPX}
	[P_-(z, \bz), X^-(z', \bz')] = - i \gamma^{z\bz} \delta^{(2)}(z - z') \,. 
\end{equation}
Note that, due to index raising and lowering conventions of lightcone coordinates, these commutation relations imply that $P^{+}$ is conjugate to $X^{-}$; the ingoing and outgoing conjugate pairs in Figure~\ref{fig:shockwave} are accordingly shown in colored pairs.

It has been previously suggested that the commutation relation \eqref{eq:hooftPX} applies near generic horizons~\cite{Verlinde:1991iu, Verlinde:2019xfb,Verlinde:2022hhs}. In this paper, we will show that \eqref{eq:hooftPX} is equivalent to the canonical commutation relations of soft modes in four-dimensional asymptotically flat spacetimes.  To establish this equivalence, we show that a linearized Bondi metric having a leading memory effect can be obtained by a diffeomorphism from a shockwave solution to the linearized Einstein's equations sourced by $T_{uu}^{M} \propto \delta(u - u_0)$. The equations of motion then imply a relation between the shockwave momentum $P^+$ and the soft graviton mode $N$, namely
\begin{equation}
\label{eq:constraint0}
	P_-(z,\bz) = \frac{1}{32 \pi G_N} \Box (\Box + 2) N(z, \bz)  \,,
\end{equation}
where $\Box$ is the transverse Laplacian (to be defined explicitly later). Physically, $N$ measures the gravitational memory effect \cite{Strominger:2014pwa}. We therefore establish a relation between the time delay acquired by a probe propagating in the shockwave background and the gravitational memory effect encapsulated concretely in the memory mode $N(z,\bz)$.\footnote{Note that our analysis focuses on a spherical shockwave background rather than a planar shockwave background, and is therefore not in contradiction with the results of \cite{Tolish:2014bka}, where the ``memory effect'' for a planar shock is shown to vanish.}
We also find that the shockwave metric enjoys a supertranslation symmetry, which allows for shifts in $X^-$ to be traded for shifts of the Goldstone mode $C$ parametrizing the memory metric, thereby allowing us to identify
\begin{equation}
	\label{X-}
	X^- = -C(z, \bz) \, .
\end{equation}
Given the identifications \eqref{eq:constraint0} and \eqref{X-}, we can immediately demonstrate that the commutation relation between soft and Goldstone operators, given by \cite{He:2014laa} 
\begin{equation}\label{eq:Poisson-CN0}
	\square_z(\square_z+2)\left[C(z,\bz), N(z',\bz') \right] = -32\pi i G_N \gamma^{z\bz} \delta^{(2)}(z - z') \,,
\end{equation}
implies the 't Hooft commutation relation \eqref{eq:hooftPX} between $P_-$ and $X^-$. Had we considered an ingoing shockwave instead of an outgoing one, we would have established the 't Hooft commutation relation between $P_+$ and $X^+$.

This equivalence has far-reaching consequences. The 't Hooft commutation relations in flat space were the foundation of a derivation in \cite{Verlinde:2022hhs}, which demonstrated that metric perturbations due to quantum effects are related to fluctuations of the modular Hamiltonian $K$ in the vacuum state of an empty causal diamond, recovering the well-known area law $\langle K \rangle = \langle \Delta K^2 \rangle = \frac{A}{4 G_N}$, where $A$ is the area of the bifurcate horizon of the causal diamond. We view our results as evidence that such modular relations can indeed be applied to light-sheet horizons in flat space, and that modular fluctuations are related to soft effects in celestial holography (related directions were explored in \cite{Kapec:2016aqd}).  These modular fluctuations, in turn, are at the center of a proposal of detectably large spacetime fluctuations in quantum gravity~\cite{Verlinde:2019xfb,Verlinde:2019ade,Banks:2021jwj}.  Thus, our work opens new directions in utilizing gravitational memory to study observational signatures of quantum gravity.

The organization of this paper is as follows. In Section~\ref{ssec:lightcone}, we give a quick overview of a spherical shockwave solution, and show how by taking an appropriate limit we recover the more familiar Aichelburg-Sexl planar shockwave solution. In Section \eqref{sec:bms}, we review the derivation of the memory effect in asymptotically flat spacetimes and its relation to asymptotic symmetries. In Section \eqref{ssec:partialBondi}, we discuss a class of Bondi metrics with relaxed fall-off conditions in the large $r$ limit. In Section~\ref{sec:memory}, we demonstrate that under a diffeomorphism, the shockwave metric considered in Section~\ref{ssec:lightcone} is transformed into a metric that agrees with the Bondi metric up to overleading contact terms. This allows us to determine in Section~\ref{ssec:new-shockwave} that the standard asymptotically flat memory metric is obtained via the same diffeomorphism from a shockwave solution to the linearized Einstein equations with a delta function perturbation turned on. The symplectic form of this new shockwave metric agrees with that of the standard memory metric subject to an identification between the shockwave and memory variables. This allows us to determine in Section~\ref{sec:hooft} how the soft graviton mode $N$ and the Goldstone mode $C$ are respectively related to the momentum $P^+$ and position $X^-$ of the outgoing shockwave and establish the 't Hooft commutation relations from the commutators of $N(z,\bz)$ and $C(z,\bz)$. Finally, we make some concluding remarks and speculate on future directions in Section~\ref{sec:conclusion}.

\section{Preliminaries}
\label{sec:prelim}

In this section, we give a brief overview of the background material necessary for our analysis. We begin by discussing some basics of shockwave metrics and reviewing the 't Hooft commutation relations in Section~\ref{ssec:lightcone}. In Section~\ref{sec:bms}, we review the relation between supertranslations and the memory effect in the Bondi gauge. Finally, in Section~\ref{ssec:partialBondi}, we introduce a class of metrics obeying slightly weaker boundary conditions compared to asymptotically flat metrics and hence containing the shockwave backgrounds.

\subsection{Shockwaves and the 't Hooft Commutation Relation} \label{ssec:lightcone}

An energetic massless particle in Minkowski spacetime sources a gravitational field referred to as an impulsive wave or a shockwave.\footnote{Impulsive waves are different from shockwaves in that they do not require a non-vanishing energy-momentum source, but the terminologies are often used interchangeably. See also \cite{Penrose:1972xrn,Freidel:2022}.} Such solutions were first found by Aichelburg and Sexl \cite{Aichelburg} and later studied by Dray and 't Hooft in the context of an ultra-boosted particle, {\it e.g.} one moving along the horizon of a black hole \cite{dray-hooft}. 

Here, we consider a shockwave metric in retarded spherical coordinates $X^\mu = (u,r,x^A)$, where $u=t-r$ and $x^A = (z,\bz)$ are the stereographic coordinates on the sphere. This describes an outward propagating spherical shock, and is given by 
\begin{align}\label{AS-met}
	ds^2 = -du^2 - 2du\,dr + 2r^2\g_{z\bz}dz\,d\bz + \underbrace{\amp(z,\bz) \delta(u-u_0)}_{h_{uu}} du^2  \,,
\end{align} 
where $\g_{z\bz} = \frac{2}{(1+z\bz)^2}$ is the metric on the unit sphere, and $D_A$ is the covariant derivative with respect to the transverse space metric $\g_{AB}$.  Note that $\alpha(z,\bar z)$ has dimensions of length. We will throughout this paper adopt the convention where lowercase Greek letters $\mu,\nu,\ldots$ are spacetime indices, and uppercase Latin letters $A,B,\ldots$ are transverse spatial indices. This metric describes a gravitational shockwave exiting future null infinity $\ci^+$ at retarded time $u = u_0.$  

To ensure that the metric \eqref{AS-met} obeys the Einstein equations, the function $\amp(z,\bz)$ is necessarily constrained. The only non-vanishing components of the Ricci tensor are 
\begin{equation}
\begin{split}
	R_{uu} &= - \frac{1}{2r^2} \Box h_{uu} - \frac{1}{r} \p_u h_{uu} \,,  \\
R_{z\bz} &= R_{\bz z} = \gamma_{z\bz} h_{uu} \,,
\end{split}
\end{equation}
where $\Box \equiv D^AD_A$ is the transverse Laplacian, and the Ricci scalar is
\begin{equation} 
R = \frac{2h_{uu}}{r^2} \,.
\end{equation}
Therefore, the Einstein equations, given by
\begin{equation} 
R_{\mu\nu} - \frac{1}{2} g_{\mu\nu} R = 8\pi G_N T_{\mu\nu}^M \,,
\end{equation}
have only two nonzero components, which are
\begin{equation} 
\label{sphere-EE}
\begin{split}
\left(-\frac{1}{2r^2} \Box - \frac{1}{r} \p_u + \frac{1}{r^2} \right) h_{uu} - \frac{h_{uu}^2}{r^2} &= 8\pi G_N T^M_{uu}\,, \\
\frac{h_{uu}}{r^2} &= 8 \pi G_N T_{ur}^M\,.
\end{split}
\end{equation}
Writing
\begin{align}\label{eq:matterT}
	T^M_{\mu\nu} = \sum_{n=2}^\infty \frac{T^{M(n)}_{\mu\nu}}{r^n}\,,
\end{align}
and substituting $h_{uu}$ from \eqref{AS-met} into \eqref{sphere-EE}, we conclude that the metric \eqref{AS-met} is a solution to the Einstein equations with a matter source 
\begin{equation}
\label{EE}
\begin{split}
	\left(\Box - 2 \right)\amp(z,\bz) \delta(u - u_0) + \mathcal{O}(\delta', \delta^2)  &= -16 \pi G_N T_{uu}^{M(2)} \,, \\
	 \amp(z,\bar{z}) \delta(u - u_0) &= 8\pi G_N T_{ur}^{M(2)} \,,
\end{split}
\end{equation}
where $\CO(\delta',\delta^2)$ indicates terms that are proportional to $\delta(u-u_0)^2$ and $\delta'(u-u_0)$. We will be mainly interested in linearized solutions, in which case quadratic terms in the delta functions can be ignored.\footnote{Alternatively, the quadratic terms correspond to a gravitational contribution to the stress tensor and may be accounted for by redefining the right-hand side of \eqref{EE}.}  As for $\delta'(u-u_0)$, we argue that such terms integrate to zero on any definite $u$ interval,\footnote{There is a subtlety if the integration endpoint involves exactly $u = u_0$, but this is a set of measure zero and we will ignore it.} and hence will not contribute to the total energy given by the $u$ integral of $T_{uu}^M$. It is straightforward to show that $\nabla^\mu T^{M}_{\mu\nu} = 0$, so the stress tensor is conserved. In Section \ref{ssec:new-shockwave} we will introduce a closely related shockwave solution to the linearized Einstein equations with vanishing $T_{ur}^{M(2)}$.

We now briefly discuss how to obtain the familiar Aichelburg-Sexl planar shock studied in \cite{Aichelburg} from the metric \eqref{AS-met}. To this end, we want to zoom into a particular patch of the celestial sphere by taking $z,\bz \ll 1$ (in this region $\g_{z\bz} \to 2$). If we then introduce the lightcone coordinates $(x^-,x^+)$ and transverse coordinate $(x_1,x_2)$, related to the retarded coordinates in \eqref{AS-met} via
\begin{align}
\label{eq:lc-coords}
	x^- = u\,, \quad x^+ = u+2r\,, \quad x_1 = r(z+\bz)\,, \quad x_2 = -ir(z-\bz)\,,
\end{align} 
the metric becomes
\begin{align}\label{eq:planar-lim}
	ds^2 = -dx^- dx^+ + dx_1^2 + dx_2^2 +  \amp\bigg( \frac{x_1}{r},\frac{x_2}{r} \bigg) \delta(x^- -x^-_0) (dx^-)^2 + \CO(z,\bz)\,,
\end{align}
where $\CO(z,\bz)$ indicates terms that are higher order in $z,\bz$ and $r = \frac{1}{2}(x^+ - x^-)$. We recognize \eqref{eq:planar-lim} as the Aichelburg-Sexl planar shock, which satisfies the Einstein equation
\begin{equation}\label{eq:planar_EE}
	 \Box_{\perp}  \amp\bigg(\frac{x_1}{r},\frac{x_2}{r}\bigg)  \delta (x^- - x^-_0) = -16\pi G_N T_{--}^M \,,
\end{equation} 
where $\Box_\perp$ is the transverse Laplacian in the $x_1,x_2$ coordinates.

We conclude this subsection with a brief review of the 't Hooft commutation relations proposed to capture quantum dynamics of particles scattering near black hole horizons. We will largely follow the presentation given in Section 11 of \cite{tHooft:1996rdg}. 
Recall the lightcone coordinates $(x^-, x^+)$, which are related to the retarded Bondi coordinates $(u, r)$ in \eqref{AS-met} by \eqref{eq:lc-coords}. In these coordinates, the shockwave metric \eqref{AS-met} takes the form 
\begin{equation}
\label{eq:shock}
	ds^2 = -dx^- dx^+ + \frac{(x^+ - x^-)^2}{2} \gamma_{z\bz} dz d\bz + \amp(z, \bz) \delta(x^- - x^-_0) (dx^-)^2 \,,
\end{equation}
which falls into the class of metrics analyzed in \cite{dray-hooft}. However, whereas the analysis in \cite{dray-hooft} was done in a Schwarzschild black hole background, we emphasize that no black hole is present in our setup. We expect our analysis to apply to  any sufficiently large causal diamonds with metrics that admit a large $r$ expansion of the form \eqref{bondi-met2} near the boundary \cite{Verlinde:2019xfb,Verlinde:2019ade, Freidel:2020xyx,Freidel:2020svx,Freidel:2020ayo,Freidel:2021cjp,Verlinde:2022hhs,Banks:2021jwj,Zurek:2022xzl}. Note that 
\begin{align}\label{pm-indices}
	x_{\pm} = -\frac{1}{2} x^\mp \,, \qquad p_\pm = -\frac{1}{2} p^\mp\,.
\end{align}

Substituting \eqref{eq:shock} into the Einstein equations, we find the $--$ component to be\footnote{We have dropped terms $\sim \delta^2, \delta'$ as in \eqref{EE}. Notice also that \eqref{boxh} superficially resembles the planar Einstein equations \eqref{eq:planar_EE}, even though we are still in spherical coordinates. Indeed, comparing with \eqref{EE}, we see that the shift to the Laplacian drops out when we write the Einstein equations in lightcone coordinates.} 
\begin{equation} 
\label{boxh}
\Box \amp(z, \bz) \delta(x^- - x^-_0) = -16\pi G_N T_{--}^{M(2)}\,.
\end{equation}
To solve for $\amp$, let us define the Green's function for the transverse Laplacian\footnote{We normalized the Green's function so that in the planar limit $\Box G(z-z') \to \delta^{(2)}(z-z')$ since $\g_{z\bz} \to 2$.}
\begin{align}\label{eq:Green_fcn}
	\Box G(z - z') = 2 \g^{z\bz} \delta^{(2)}(z - z') \, \implies G(z - z') = \frac{1}{2\pi} \log|z - z'|^2 \,.
\end{align}
It is clear then from \eqref{boxh} that a generic matter distribution in the transverse directions with stress tensor 
\begin{align}\label{eq:gen-T}
	T^{M(2)}_{--}(x^-,z,\bz) = p_-(z,\bz)\delta(x^--x_0^-)
\end{align}
leads to a shockwave profile
\begin{align}\label{eq:h-green}
\begin{split}
	\amp(z,\bz) &= - 8\pi G_N \int d^2z' \,\gamma_{z'\bz'} \, G(z-z')p_-(z',\bz') \\
	&= 4\pi G_N \int d^2z' \, \gamma_{z'\bz'}\, G(z-z')p^+(z',\bz')\,,
\end{split}
\end{align}
where we used \eqref{pm-indices} to switch $p_-$ to $p^+$. For a single source particle of constant outgoing momentum $p_-$, we have $p_-(z, \bz) = p_- \delta^{(2)}(z - z_0)$.

In the presence of the shock \eqref{eq:h-green}, a probe particle propagating along the future horizon (or equivalently along $x^-$) will acquire a time delay $\delta x^+$ for $x^- > x_0^-$:
\begin{align} 
	x^+\left(x^- > x_0^-, z, \bz\right) = x^+\left(x^- < x_0^-, z, \bz\right) + \delta x^+ \,.
\end{align}
This setup is illustrated in Figure \ref{fig:shockwave}.
The magnitude of the time delay is determined by the wave equation in the shockwave background \eqref{eq:shock}, and is given by \cite{tHooft:1987vrq} 
\begin{align}
\label{eq:time-delay}
	\delta x^+ = \amp(z,\bz)  \, .
\end{align}
The propagation of this particle in the shockwave background is therefore characterized by the phase shift
 \begin{align}
 \label{S-matrix}
 \begin{split}
 \langle \rm out| \rm in \rangle &\propto \exp{\left\{i \int d^2 z \,\gamma_{z\bz} \, \delta x^+(z, \bz) p_+(z, \bz) \right\}} \\
&=  \exp{\left\{- 2\pi iG_N \int d^2 z\, \gamma_{z\bz} \int d^2z' \,\gamma_{z'\bz'}\, p^+(z',\bz') G(z - z') p^-(z,\bz)\right\}} \,,
 \end{split}
 \end{align}
 where in the second line we substituted $\delta x^+$ in terms of $p^+(z, \bz)$ according to \eqref{eq:h-green} and \eqref{eq:time-delay}. 

Thus far, all the equations we derived above follow from the classical equations of motion. By \eqref{eq:time-delay}, we have a relation between $\delta x^+$ and $p^+$, and there is no classical shift of the shockwave position $x^-$. However,
in the special case where the shockwave is sourced by a point particle, \eqref{S-matrix} may be regarded as the high-energy limit of a 2-to-2 scattering amplitude \cite{tHooft:1987vrq, Amati:1992zb, Giddings:2007bw, Giddings:2011xs}. As such, it is clear that one would arrive at the same result \eqref{S-matrix} by changing reference frame, or equivalently, by exchanging the roles of the source and probe particles. The metric in this case takes the same form as in \eqref{eq:shock} with $x^+ \leftrightarrow x^-$, namely a shock located at $x^+ = x^+_0$, and the propagating particle in this new background will acquire a shift
 \begin{equation}\label{eq:xm-shift}
	\delta x^-(z) = 4\pi G_N \int d^2 z' \, \gamma_{z'\bz'}\, G(z - z') p^-(z')\,,
\end{equation}
paralleling \eqref{eq:h-green}.

This motivated 't Hooft to promote $p^{\pm},~\delta x^{\pm}$  to operators $P^{\pm},~X^{\pm}$ and postulate the commutation relations 
\begin{equation} 
\label{eq:commutators}
\begin{split}
	[P^\pm(z), X_\pm(z')] &= - i \g^{z\bz} \delta^{(2)}(z - z')\,, \qquad [P^{\pm}(z), P^{\pm}(z')] = [X_{\pm}(z), X_{\pm}(z')] = 0 \,,
\end{split}
\end{equation}
where we have suppressed the dependence of all operators on $\bz$ for clarity. These commutation relations can then be used to recover the phase shift \eqref{S-matrix}, which we review in Appendix~\ref{app:S-matrix}. Remarkably, we will later show in Section~\ref{sec:hooft} that $P_- = -\frac{1}{2}P^+$ is directly related to the leading gravitational memory operator \cite{Strominger:2014pwa, Compere:2019odm, Freidel:2022skz}, while $X^- = -2X_+$ will be identified with the Goldstone operator involved in discussions of infrared divergences \cite{Choi:2017bna, Donnay:2018neh, Arkani-Hamed:2020gyp, Nguyen:2021ydb, Kapec:2021eug, Kapec:2022hih}. 

\subsection{Asymptotically Flat Metrics and Gravitational Memory}\label{sec:bms}

In this subsection, we review the standard framework used to describe gravitational radiation from isolated sources in spacetime \cite{Strominger:2014pwa}. We will assume that observers are located at a relative distance much larger than the extent of the sources, where spacetime is approximated by Minkowski space with metric $g^{(0)}_{\mu\nu}$. Gravitational radiation is characterized by linear perturbations $h_{\mu\nu}$ around this flat background, so that 
\begin{equation}
	g_{\mu\nu} = g^{(0)}_{\mu\nu} + h_{\mu\nu}\,.
\end{equation}
As appropriate for the propagation of outgoing gravitational waves, we will as in the shockwave case adopt retarded spherical coordinates $X^{\mu} = (u, r, x^A)$. A convenient gauge that the metric can be expressed in is the Bondi gauge \cite{Bondi:1962px, Sachs:1962zza}, which is defined to obey the condition
\begin{equation} 
\label{eq:Bondi-gauge}
	g^{\mu\nu} \p_{\mu} u \, \p_{\nu} u = g^{\mu\nu} \p_{\mu} u \,\p_{\nu} x^A = 0 \quad\implies\quad g_{rr} = g_{rA} = 0\,,
\end{equation}
as well as the determinant condition \cite{Barnich:2010eb}
\begin{equation}
\label{eq:det}
	\p_r \,{\rm det} \bigg( \frac{ g_{AB}}{r^2} \bigg)  = 0\,.
\end{equation}
Physically these correspond to requiring the normal vectors $\p_{\mu} u$ to be null and the angular coordinates to be constant along null rays, or equivalently that gravitational waves propagate along the radial direction and have spherical wavefronts. 

At large $r$, the perturbations $h_{\mu\nu}$ admit expansions of the form\footnote{This expansion may include additional logarithmic terms  \cite{Winicour, Chrusciel:1993hx, Geiller:2022vto}, which will not play a role here.}
\begin{equation}
\label{eq:exp}
h_{\mu\nu} = \sum_{n = -1}^{\infty} \frac{h_{\mu\nu}^{(n)}}{r^n} \, ,
\end{equation}
with the fall-offs of each component determined by the choice of boundary conditions. Solving the asymptotic radial Einstein equations subject to the standard fall-offs proposed in \cite{Bondi:1962px,Sachs:1962zza} (for a review see \cite{Strominger:2017zoo} and references therein) leads to the metric for an asymptotically flat spacetime near future null infinity $\ci^+$
\begin{align}\label{bondi-met2}
\begin{split}
	ds^2 &= -du^2 - 2 du\,dr + 2 r^2\g_{z\bz} dz\,d\bz \\
	&\qquad + \frac{2m_B(u,z,\bz)}{r}du^2 + \big( rC_{zz}(u,z,\bz) dz^2 + \cc \big) + \big( D^z C_{zz}(u,z,\bz) du\,dz + \cc \big) + \cdots\,.
\end{split}
\end{align}
Here $m_B$ is the Bondi mass aspect (note that, despite its label as a mass, it has dimensions of length, effectively having absorbed a factor of $G_N$), and $\cdots$ denote subleading corrections in the large $r$ expansion. The gravitational radiation is characterized by the news tensor, related to the shear $C_{zz}$ by
\begin{align}\label{Ndef}
	N_{zz}(u,z,\bz) = \p_u C_{zz}(u,z,\bz)\,.
\end{align}
Therefore, spacetimes that can be written in the form of \eqref{bondi-met2} are often referred to as radiative spacetimes, and if $N_{zz} = 0$ then the metric describes a vacuum spacetime. Note from \eqref{Ndef} that $N_{zz} = 0$ does not imply the vanishing of $C_{zz}$. In particular, imposing $N_{zz} = m_B = 0$, we obtain from the $uz$ component of the vacuum Einstein equations that 
\begin{equation}
\label{eq:vacua} 
C_{zz}^{\rm vac} = - 2 D_z^2 C(z, \bz) \,.
\end{equation}
As such, vacuum spacetimes are parametrized by an arbitrary function $C(z,\bz)$ on the sphere. This variable is known as the Goldstone mode canonically conjugate to the zero mode of the news tensor or the soft graviton mode \cite{Strominger:2017zoo}, as we will review below. Moreover, these different vacua are related by large diffeomorphisms called supertranslations.

Further imposing the constraint equations, broadly defined to be all the remaining Einstein equations, order by order in a large $r$ expansion provides relations between $m_{B}$ and $C_{zz}$, and also among further subleading metric components. For example, at $\mathcal{O}(r^{-2})$ the $uu$ component of the Einstein equations takes the form
\begin{equation}
\label{eq:uu-constraint}
	\p_u m_B = \frac{1}{4}\left(D_z^2 N^{zz} + D_{\bz}^2 N^{\bz\bz} \right) - T_{uu}\,, 
\end{equation}
where
the stress tensor $T_{uu}$ (by convention {\it dimensionless} here) is defined to be
\begin{equation}\label{Tuu-def}
	T_{uu} \equiv \frac{1}{4} N_{zz}N^{zz}  + 4\pi G_N T_{uu}^{M(2)}\,, 
\end{equation}
with $T_{uu}^{M(2)}$ being the coefficient of the $r^{-2}$ component of the matter stress tensor defined in \eqref{eq:matterT}.
Consider now the $\ell = 0,1$ spherical harmonic modes of $N_{zz}$. Recalling $Y_{\ell=0} \propto 1$ and $Y_{\ell=1} \propto \left( \frac{\bz}{1+z\bz}, \frac{1-z\bz}{1+z\bz}, \frac{z}{1+z\bz}\right)$, it is easy to see that 
\begin{align}
	D_A^2 Y_{\ell=0,1}  = 0\,.
\end{align}
Thus, the linear terms in $N^{zz}$ and $N^{\bz\bz}$ drop out for the $\ell=0,1$ modes, which means $\p_u m_B$ cannot generically vanish unless $T_{uu}^M$ fails to satisfy positive energy conditions in the $\ell=0,1$ modes. Nevertheless, we will henceforth assume this, and leave the inclusion of a nontrivial Bondi mass for future work. 

The asymptotic symmetries of asymptotically flat spacetimes are the transformations that preserve the form of the metric \eqref{bondi-met2} and are known as the BMS symmetries \cite{Bondi:1962px, Sachs:1962zza}. A particular subset of BMS symmetries are the supertranslations, which are parametrized by a function $f(z,\bz)$ on the celestial sphere and are generated by the vector fields
\begin{align}\label{xi_super}
	\xi = f\p_u + D^zD_z f \p_r - \frac{1}{r}  D^A f \p_A  \,.
\end{align}
Under a diffeomorphism parametrized by vector $\xi$, the coordinates and metric transform according to 
\begin{align}\label{diffeo_vec}
	\delta x^\mu = - \xi^\mu \,, \qquad \delta g_{\mu\nu} = \CL_\xi g_{\mu\nu} = 2\nabla_{(\mu} \xi_{\nu)}\,,
\end{align}
where we are using $\nabla_\mu$ as opposed to $D_A$ for spacetime covariant derivatives, and symmetrizing indices by the convention $v_{(a}w_{b)} = \frac{1}{2}(v_a w_b + v_b w_a)$. Explicitly, under a supertranslation the coordinates shift via
\begin{align}\label{supertranslation}
	u \to u - f\,, \quad r \to r - D^zD_z f\,, \quad x^A \to x^A + \frac{1}{r} D^A f\,,
\end{align}
and the shear $C_{zz}$ and news $N_{zz}$ transform as \cite{Barnich:2011mi, Strominger:2017zoo} 
\begin{align}\label{supertranslation-C}
	\CL_\xi C_{zz} = f N_{zz} - 2 D_z^2 f\,, \qquad \CL_\xi N_{zz} = f \p_u N_{zz}\,.
\end{align}
The first equation in \eqref{supertranslation-C} implies that supertranslations act non-trivially on vacua parametrized by \eqref{eq:vacua}, namely
 \begin{equation}
 \label{eq:Goldstone}
 C(z, \bz) \rightarrow C(z, \bz) + f(z, \bz) \,.
 \end{equation}

Gravitational flux characterized by $N_{zz} \neq 0$ for $u \in [u_i, u_f]$ generally induces transitions between the early and late vacua at $u < u_i$ and $u > u_f$. In particular, because this is no longer a vacuum spacetime, we do not expect \eqref{eq:vacua} to hold, and the shear $C_{zz}$ is now a function of $u$. A sharp probe of this is the gravitational memory effect \cite{zeldovich,Braginsky:1985vlg,thorne, Ludvigsen:1989kg, Christodoulou:1991cr, Wiseman:1991ss, PhysRevD.45.520, Strominger:2014pwa}, where gravitational radiation induces a net relative transverse displacement in the trajectories of nearby detectors or probe particles. This displacement is directly related to the change in the shear between times $u_i$ and $u_f$ before and after the gravitational pulse, given by
\begin{equation}
\label{eq:delta-C}
	\Delta C_{zz} \equiv C_{zz}(u_f = \infty ,z, \bz) - C_{zz}(u_i = -\infty, z, \bz)  = \int_{-\infty}^{\infty}du\, N_{zz}(u,z,\bz) \equiv D_z^2 N(z,\bz) \,,
\end{equation}
where we can view the last equality as the definition of the field $N(z,\bz)$. Note that $N(z,\bz)$ is directly related to the zero mode of the news and can be shown to correspond to the leading soft graviton mode upon quantization \cite{He:2014laa}. We can determine $N(z,\bz)$ by integrating the constraint \eqref{eq:uu-constraint}, so that  
\begin{equation}
\label{eq:constraint-memory}
\frac{(\gamma^{z\bz})^2}{4}\left(D_z^2D_{\bz}^2 + D_{\bz}^2 D_{z}^2 \right)N(z, \bz) = \Delta m_{B} + \int_{-\infty}^{\infty} du \,T_{uu} \,.
\end{equation}
For the special case of a delta function localized stress tensor $T_{uu} \propto\delta(u - u_0)$ and $m_B = 0$,  we have $N_{zz} \propto \delta(u-u_0)$ by \eqref{eq:uu-constraint}, which in turn implies by applying \eqref{eq:delta-C} over an integration from $-\infty$ to $u$ that 
\begin{equation} 
\label{eq:Czz-step}
C_{zz}(u,z,\bz) =  D_{z}^2 N(z, \bz) \Theta(u - u_0) - 2D^2_z C(z,\bz) \,,
\end{equation}
where $\Th(x)$ is the unit step function. Here $N(z, \bz)$ is determined in terms of the flux by \eqref{eq:constraint-memory}, and $C(z,\bz)$ can be viewed as an initial condition, related to the vacuum $C_{zz}^{\rm vac}$ by \eqref{eq:vacua}. We emphasize that, according to \eqref{eq:Goldstone}, supertranslations induce shifts in $C(z,\bz)$ while leaving $N(z,\bz)$ invariant. On the other hand, the vacuum transition between early and late times induced by flux is itself parametrized by a supertranslation in the sense that
\begin{equation} 
C_{zz}(u_f,z, \bz) = C_{zz}(u_i, z, \bz) + D_z^2 N(z, \bz) \,.
\end{equation}
Consequently, one can regard supertranslations as relating physically inequivalent configurations \cite{Strominger:2014pwa}.

Solutions analogous to \eqref{eq:Czz-step} in QED were discussed in \cite{Kapec:2017tkm}. Choosing $C(z,\bz)$ and $N(z,\bz)$ such that they diagonalize boosts towards a point $(z_0, \bz_0)$ on the celestial sphere yields solutions related to the conformally soft wavefunctions in celestial holography \cite{Donnay:2018neh,Pasterski:2021fjn,Pasterski:2021dqe}. As argued in \cite{He:2014laa}, the phase space of gravity in asymptotically flat spacetimes has to be augmented by $N(z,\bz)$ and $C(z, \bz)$.  Upon quantization, the former correspond to insertions of soft gravitons while the latter are key constituents of the gravitational dressings \cite{Arkani-Hamed:2020gyp, Nguyen:2021ydb, Kapec:2021eug, Kapec:2022hih}. 
As we review below, $N(z,\bz)$ and $C(z, \bz)$ are also canonically conjugate to one another \cite{He:2014laa, Donnay:2018neh}. Consequently, the role of the gravitational dressings is to supply the asymptotic states with the amount of soft charge (related to soft gravitons) necessary to ensure that the net large gauge charge is conserved in any scattering process \cite{Kapec:2017tkm, Choi:2017bna}. 

The commutator between the fields $C(z,\bz)$ and $N(z,\bz)$ was first worked out in \cite{He:2014laa} by starting from the canonical commutation relations in gravity given by (also see Appendix~\ref{app:symplectic}) \cite{Ashtekar:1978zz, Ashtekar:1981sf}
\begin{equation}\label{eq:Poisson-CN}
	\big\{C_{\bz\bz}(u,z,\bz), N_{ww}(u',w,\bw) \big\} = 16\pi G_N \gamma_{z\bz}\delta(u - u') \delta^{(2)}(z - w) \,,
\end{equation}
and then imposing suitable boundary conditions. Note that from  \eqref{eq:vacua} and \eqref{eq:Czz-step}, it is clear that 
\begin{align}\label{eq:Czz-bdy}
	\lim_{u \rightarrow -\infty} C_{zz}(u,z, \bz) = - 2D_z^2 C(z, \bz) = C_{zz}^{\rm vac}(z,\bz) \,.
\end{align}
As a result (quantum commutators $[\cdot,\cdot]$ are identified with the Dirac brackets $i\{\cdot,\cdot\}$),
\begin{align}\label{CN-com}
\begin{split}
	[N(w,\bw),C(z,\bz)] &= 8i G_N S \log|z-w|^2 \,,
\end{split}
\end{align}
where $S = \frac{(z-w)(\bz-\bw)}{(1+z\bz)(1+w\bw)}$ and we used the identities
\begin{align}\label{S-identity}
\begin{split}
	D_w^2 \big( S \log|z-w|^2 \big) &= \frac{S}{(z-w)^2} \,, \\
	D_\bz^2D_w^2\big( S \log|z-w|^2 \big) &= \pi\g_{z\bz}\delta^2(z-w) \,.
\end{split}
\end{align}
In the planar limit, the sphere is flattened to a plane ($\gamma_{z\bz} \rightarrow 2$) and $S \rightarrow |z - w|^2$. Consequently, \eqref{CN-com} simply becomes 
\begin{equation} 
[N(w,\bw),C(z,\bz)] = 8i G_N |z - w|^2 \log|z-w|^2\,.
\end{equation}

\subsection{Relaxing the Boundary Conditions}\label{ssec:partialBondi}

In this short subsection, we introduce a class of metrics obeying \eqref{eq:Bondi-gauge} and \eqref{eq:det}, but with $h_{uu}$ allowed to be finite in the large $r$ limit. Notice that this class of metrics contains the shockwave metric \eqref{AS-met}, and it is more general than metrics obeying the fall-off conditions \eqref{bondi-met2}, where $h_{uu} \sim \CO(r^{-1})$. These generalized boundary conditions, as well as a partial gauge fixing where \eqref{eq:det} is not imposed and \eqref{eq:exp} is allowed to include logarithmic corrections, were extensively studied in \cite{Geiller:2022vto}. 

Metrics obeying \eqref{eq:Bondi-gauge} take the form
\begin{equation} 
\label{eq:partial-bondi}
ds^2 = e^{2\beta} \frac{V}{r} du^2 - 2 e^{2\beta} du dr + g_{AB} \left(dx^A - U^A du \right) \left(dx^B - U^B du \right) \,,
\end{equation}
where in particular $V,$ $U^A$ and $\beta$ are allowed to obey weaker fall-offs at large $r$ than those specified by \eqref{bondi-met2} in a way consistent with the Einstein equations. Before imposing the determinant condition \eqref{eq:det}, such metrics enjoy an enhanced diffeomorphism invariance parametrized by
\begin{equation} 
\label{eq:diffeo}
\xi = F\p_u + \xi^r \p_r - \frac{1}{r} D^A F \p_A  + \cdots \,, 
\end{equation}
where $F = F(u, z, \bz)$ and $\xi^r$ is an arbitrary function of $(u,r,z,\bz)$, and $\cdots$ indicate subleading terms in the large $r$ expansion. 

Imposing the determinant condition \eqref{eq:det} fixes $\xi^r$ in terms of $F$, namely
\begin{equation}
\label{eq:det-fix}
\begin{split}
	\mathcal{L}_{\xi} g_{z\bz} = 0 \quad\implies\quad &   r^2 D_z \xi_{\bz} + r^2 D_{\bz} \xi_{z} + 2r \gamma_{z\bz} \xi^r = 0 \\
	\implies\quad & \xi^r = -\frac{r}{2} D_A \xi^A = \frac{1}{2}\Box F \,.
\end{split}
\end{equation}
Substituting this back into \eqref{eq:diffeo}, we see that a generic diffeomorphism preserving \eqref{eq:partial-bondi} is parametrized by a single function $F$, given by
\begin{align}\label{eq:diffeo2}
	\xi = F\p_u + \frac{1}{2} \Box F\p_r - \frac{1}{r} D^A F \p_A + \cdots\,.
\end{align}
Notice that we recover \eqref{supertranslation}, but with $f$ promoted to a function $F$ that may depend on $u$, due to not having demanded the vanishing of $\CO(r^0)$ component of  $h_{uu}$. In this case, the vector fields \eqref{eq:diffeo2} have non-vanishing divergence $\nabla \cdot \xi = \p_u F \neq 0$, which implies that the associated diffeomorphisms will in general change the trace $h = h_\mu{}^{\mu}$. The diffeomorphism considered in the next section falls into this class.\footnote{Alternatively, one can demand the trace to be preserved, and this forces one to relax the determinant condition \eqref{eq:det-fix}, which was the route taken in \cite{Geiller:2022vto}.} 

\section{Relating Shockwaves to Gravitational Memory}
\label{sec:memory}

In this section we establish an equivalence on the future null horizon away from $u = u_0$ between a shockwave metric closely related to \eqref{AS-met} and Bondi metrics of the form \eqref{bondi-met2} with $C_{zz}$ given by \eqref{eq:Czz-step}. In particular, we first demonstrate in Section~\ref{ssec:diffeo} that there exists a diffeomorphism parametrized by \eqref{eq:diffeo2} that relates the shockwave metric \eqref{AS-met} to a metric that resembles \eqref{bondi-met2} up to contact terms. This diffeomorphism provides a relation between $N(z,\bz)$ in \eqref{eq:Czz-step} and the shock profile $\amp(z,\bz)$. Both metrics are shown in Appendix \ref{app:symplectic} to have vanishing symplectic form, suggesting the diffemorphism is ``small.'' In Section~\ref{ssec:new-shockwave}, the same diffeomorphism is shown to relate another shockwave metric to the Bondi metric \eqref{bondi-met2}. Both metrics now have non-vanishing symplectic forms that agree, subject to the identification between the shockwave and memory variables $\alpha$ and $N$. In Section~\ref{sec:hooft} we show that the canonical commutation relation between the Goldstone and soft graviton modes $C$ and $N$ implies the 't Hooft commutation relation \eqref{eq:commutators}. This extends the analysis of 't Hooft near a Schwarzschild horizon to null horizons at large $r$.

\subsection{The Diffeomorphism} \label{ssec:diffeo}

In this subsection, we show that \eqref{AS-met} is diffeomorphic to a metric that satisfies the gauge conditions \eqref{eq:Bondi-gauge} and \eqref{eq:det} and where the shear takes the form \eqref{eq:Czz-step}. Recall that under a diffeomorphism parametrized by vector field $\xi$, the metric transforms linearly via \eqref{diffeo_vec}. Requiring that this transformation eliminates the $h_{uu}$ component of the metric \eqref{AS-met}, we find
\begin{align}
\label{solution}
\begin{split}
	0 = h_{uu} + 2 \nabla_u \xi_u  \quad \implies\quad & \p_u \xi_u = -\frac{1}{2}\amp(z,\bz) \delta(u-u_0)  \\
	\implies \quad & \xi_u = -\frac{1}{2}\amp(z,\bz)  \Th(u-u_0) + g(z,\bz)\,,
\end{split}
\end{align}
where $g(z,\bz)$ is an arbitrary function on the sphere. It is easy to see that
\begin{align}\label{eq:xi-indices}
	\xi_u = -\xi^u - \xi^r\,, \qquad \xi_r = -\xi^u\,, \qquad \xi_z = r^2\g_{z\bz} \xi^\bz\,.
\end{align}
Together with the requirement \eqref{eq:diffeo2} that Bondi gauge is preserved under the diffeomorphism, $F(u,z,\bz)$ is determined in terms of the shock profile $\amp(z,\bz)$ and $g(z,\bz)$ to be
\begin{align}\label{eq:F}
\begin{split}
	& -\xi^u - \xi^r = -\frac{1}{2}\amp(z,\bz)\Th(u-u_0) + g(z,\bz) \\
	\implies\quad & (\Box + 2) F(u,z,\bz) = \amp(z,\bz)\Th(u-u_0) - 2 g(z,\bz) \,.
\end{split}
\end{align}

We show in Appendix~\ref{app:diffeo} that diffeomorphisms parametrized by \eqref{eq:diffeo2} 
\begin{equation}
\label{eq:NC-diffeo}
	\xi = F\,\p_u + \frac{1}{2}\Box F\,\p_r - \frac{1}{r}D^A F\,\p_A \,,
\end{equation}
with $F(u, z, \bz)$ defined by \eqref{eq:F}, transform the shockwave metric \eqref{AS-met} to a form closely related to the Bondi metric introduced in \eqref{bondi-met2}, namely
\begin{align}\label{eq:partialbondi}
\begin{split}
	ds^2 &= -du^2 - 2 \,du\,dr + 2r^2\gamma_{z\bz}\,dz\,d\bz   \\
	& \qquad + \big( r C_{zz}(u,z,\bz)\,dz^2  +   D^zC_{zz}(u,z,\bz)  \,du\,dz + \cc \big)  \\
	&\qquad + N(z,\bz)\delta(u-u_0)\,du\,dr +  \big( r\p_z N(z,\bz)\delta(u-u_0)\,du\,dz + \cc \big) \,.
\end{split}
\end{align}
Here  $C_{zz}$ takes the form \eqref{eq:Czz-step}, and from \eqref{eq:app-hg} we see that $N(z, \bz)$ is determined by the shockwave background to be 
\begin{align}\label{eq:hg0}
	 -\frac{1}{2}(\Box+2) N(z,\bz) = \amp(z, \bz) \,,
\end{align}
and $C(z,\bz)$ is related to the integration constant $g(z,\bz)$ of the vector field in \eqref{solution} via
\begin{equation}\label{eq:hg1}
	- \frac{1}{2}(\Box+2) C(z,\bz) = g(z,\bz)\,.
\end{equation}
Notice that \eqref{eq:partialbondi} differs from the usual Bondi metric \eqref{bondi-met2} by terms proportional to delta functions, given in the last line of \eqref{eq:partialbondi}. As described in Section~\ref{ssec:partialBondi}, \eqref{eq:NC-diffeo} transforms the shockwave metric \eqref{AS-met}, which is traceless, to one with a nonzero trace, specifically $h_{ur} \neq 0$. The $\mathcal{O}(r)$ component of $h_{uz}$ is then necessarily nonzero to ensure that the Einstein equations are obeyed.

We conclude this subsection with a comment on the physical interpretation of diffeomorphisms parametrized by \eqref{eq:NC-diffeo}. Both the time-dependent and the time-independent components of the vector fields preserve the relaxed Bondi boundary conditions where $h_{uu} = \mathcal{O}(1)$. The time-independent component preserves the asymptotically flat large-$r$ fall-offs in \eqref{bondi-met2} while modifying the boundary value of the vacuum shear \eqref{eq:vacua}, and is considered a large gauge transformation. As such, for $g \neq 0$, \eqref{AS-met} and \eqref{eq:partialbondi} describe different theories. One could arrive at \eqref{eq:partialbondi} without performing a large gauge transformation \eqref{eq:hg1} by considering a shockwave metric in an arbitrary Minkowski vacuum. This can be achieved by turning on $C_{zz}^{\rm vac}$ in \eqref{AS-met}, as we shall see in Section~\ref{ssec:new-shockwave}.

On the other hand, the time-dependent component violates the asymptotically flat boundary conditions at large $r$ and is therefore not typically included in the asymptotic symmetry group of asymptotically flat spacetimes. Analogous gauge transformations are known to trade Coulombic for soft/memory degrees of freedom in gauge theories \cite{Verlinde:1993te, Iancu:2003xm, Pate:2019mfs}, where they were classified as residual gauge transformations. Indeed, we show in Appendix \ref{app:symplectic} that the symplectic form of \eqref{eq:partialbondi} vanishes subject to \eqref{eq:hg0}, and hence remains invariant under such time-dependent diffeomorphisms. We leave a complete understanding of time-dependent diffeomorphisms and their associated charges (if any) to future work.

\subsection{Modifying the Shockwave Metric} \label{ssec:new-shockwave}

The analysis in the previous section allows us to easily connect the standard asymptotically flat memory metric \eqref{bondi-met2} with $m_B = 0$ to a shockwave metric closely related to \eqref{AS-met}. This can be achieved by subtracting the delta function contact terms from \eqref{eq:partialbondi}, which leads us to consider a shockwave metric of the form
\begin{align}\label{AS-met-modified}
\begin{split}
	ds^2 &= -du^2 - 2\,du\,dr + 2r^2\g_{z\bz}\,dz\,d\bz + \amp(z,\bz) \delta(u-u_0) du^2 \\
	&\qquad - N(z,\bz)\delta(u-u_0)\,du\,dr - \big( r\p_z N(z,\bz) \delta(u-u_0)\,du\,dz + \cc \big) \\
	&\qquad + \big( r C_{zz}^\vac(z,\bz)\,dz^2 + D^z C_{zz}^\vac(z,\bz)\,du\,dz + \cc \big) \,,
\end{split}
\end{align}
where $C_{zz}^\vac = -2D_z^2 C$ is defined in \eqref{eq:vacua}. The first line is precisely the spherical shockwave metric \eqref{AS-met}. Note that \eqref{AS-met-modified} differs from \eqref{AS-met} both by the choice of Minkowski vacuum and by having a non-vanishing trace $h_{ur} \neq 0$. Allowing the trace to be non-vanishing implies the Einstein's equations are obeyed without a $ur$-component of the matter stress tensor (see \eqref{eq:new-eom} below). The trace will be removed by the diffeomorphism generated by \eqref{eq:NC-diffeo} with $F$ given by \eqref{eq:F}. 

Indeed, because the diffeomorphism generated by the vector field $\xi$ \eqref{eq:NC-diffeo} is linear, the metric we obtain after applying the (linearized) diffeomorphism parametrized by $\xi$ generating a small gauge transformation is precisely
\begin{align}\label{eq:bondigaugefinal}
\begin{split}
	ds^2 &= -du^2 - 2 \,du\,dr + 2r^2\gamma_{z\bz}\,dz\,d\bz   \\
	& \qquad + \big( r C_{zz}(u,z,\bz) \,dz^2  +   D^z C_{zz}(u,z,\bz) \,du\,dz + \cc \big) \,,
\end{split}
\end{align}
where $C_{zz}$ is defined in \eqref{eq:Czz-step}, and we used the identifications \eqref{eq:hg0} and \eqref{eq:hg1} with $g(z,\bz) = 0$ to ensure that $\xi$ is generating a small gauge transformation. We refer to this metric as a memory metric, which are Bondi metrics with nonzero memory mode $N(z,\bz)$. It is clear this metric is exactly of the form \eqref{bondi-met2}, which means the commutation relation \eqref{CN-com} holds.

Evaluating Einstein's equations for \eqref{AS-met-modified} and using \eqref{eq:hg0}, the only nontrivial equations are
\begin{align}\label{eq:new-eom}
\begin{split}
	\Box \amp(z,\bz)\delta(u-u_0)  + \CO(\delta') &= -16 \pi G_N T_{uu}^{M (2)} \,, \qquad  T_{ur}^{M(2)} = 0  \,.	
\end{split}
\end{align}
Thus we see that as in 't Hooft's original planar shockwave analysis, the $ur$-component of Einstein's equation vanishes. Furthermore, integrating the $uu$-component of Einstein's equation yields
\begin{align}\label{eq:new-P-}
	P_-(z, \bz) \equiv \int_{-\infty}^{u} du'\, T_{uu}^{M(2)}(u',z,\bz) = - \frac{1}{16\pi G_N} \Box \amp(z,\bz) \,,
\end{align}
for $u > u_0$. This is precisely the momentum operator defined in 't Hooft's original analysis, given in \eqref{eq:hooftP}.

\subsection{From Soft to 't Hooft Commutation Relations} \label{sec:hooft}

In this subsection, we demonstrate that \eqref{eq:commutators} is in fact implied by the canonical commutation relations \eqref{CN-com} obeyed by $C(z, \bz)$ and $N(z, \bz)$. Since the metric \eqref{AS-met-modified} is diffeomorphic to the Bondi metric \eqref{bondi-met2}, the canonical commutation relations \eqref{CN-com} remain unchanged.

First, note that by substituting \eqref{eq:hg0} into \eqref{eq:new-P-}, we are able to immediately express $P_-$ in terms of $N(z,\bz)$ for $x^- > x_0^-$, namely 
\begin{equation}
\label{eq:PN}
	P_-(z,\bz) = \frac{1}{32 \pi G_N} \Box (\Box + 2) N(z, \bz) \,.
\end{equation}
This result is not as surprising as it may seem. Under our diffeomorphism, the stress tensor $T_{--}^M$ remains the same to linear order, while the Coulombic degrees of freedom of the shockwave metric are effectively shifted into radiative ones, as one can see by comparing the  $--$ components of the Einstein equations \eqref{boxh} and \eqref{eq:constraint-memory}. As such, the relation between $P_-$ and $N$ in \eqref{eq:PN} could have been directly deduced from the equations of motion before and after the diffeomorphism. A completely analogous relation between the Coulombic and memory degrees of freedom in the high-energy limit of QCD was pointed out in \cite{Ball:2018prg}. 

To identify the shockwave variable associated to the Goldstone $C(z,\bz)$, note that diffeomorphisms with non-vanishing $g(z,\bz)$ in \eqref{eq:hg1}, and hence $C(z, \bz)$, change the boundary vacuum configuration. The metric \eqref{AS-met-modified} generalizes the shockwave metrics considered by 't Hooft by allowing for $C_{zz}^{\rm vac} \neq 0$. However, one can revert to the configuration originally considered by 't Hooft with $C_{zz}^{\rm vac} = 0$ by a supertranslation  \eqref{eq:diffeo} with $F = C(z, \bz)$ acting on coordinates while keeping the metric fixed.\footnote{Note that the metric is left invariant by both shifts of coordinates and the metric according to \eqref{diffeo_vec}.} In particular, recalling from \eqref{eq:lc-coords} that $x^- = u$, by \eqref{supertranslation} $x^-$ transforms by the simple shift
\begin{equation}
	x^{-} \rightarrow x^{-} - C\,.
\end{equation}
Promoting this field-dependent coordinate shift to an operator, we are led to write $\delta x^- = X^-$, suggesting the identification of the 't Hooft variable $X^-$ with the Goldstone mode $C$ via
\begin{equation} 
	X^-(z,\bz) = -C(z, \bz) \,,
\end{equation}
up to a field-independent variable. Note that the relative sign in this identification is consistent with the action of supertranslations \eqref{supertranslation} and \eqref{eq:Goldstone} on $X^-$ and $C$, respectively. 

Finally, since $N$ is proportional to the transverse momentum mode $P_-$ via \eqref{eq:PN}, we see that the commutator between $C$ and $N$ \eqref{CN-com} implies a nontrivial commutation relation between $P_-$ and $X^-$, which for $x^- > x^-_0$ is given by
\begin{equation}\label{eq:hooft_final}
\begin{split}
	[P_-(z,\bz), X^-(z',\bz') ] &= -\frac{1}{32\pi G_N} \Box_z (\Box_z+2) \big[ N(z,\bz), C(z',\bz') \big] \\
	&= - \frac{i}{4\pi} \Box_z (\Box_z+2) \big( S\log|z-z'|^2 \big) \\
	&= -i\gamma^{z\bz}\delta^{(2)}(z-z')\,.
\end{split}
\end{equation}
This exactly reproduces the 't Hooft commutation relation \eqref{eq:commutators}. We could have obtained the 't Hooft commutation relation involving $P_+$ and $X^+$ had we begun with an ingoing shockwave localized at $x^+ = x^+_0$ rather than an outgoing shockwave.

We can recast the commutation relation \eqref{eq:hooft_final} into a more symmetric form. Recalling the classical equations of motion \eqref{eq:h-green} and \eqref{eq:time-delay} and promoting them to operator equations, we have
\begin{align}
\begin{split}
	X^+(z,\bz) &= 4\pi G_N \int d^2z'\, \g_{z'\bz'} \, G(z-z')P^+(z',\bz') \\
	&= -8 \pi G_N \int d^2z'\, \g_{z'\bz'} \, G(z-z')P_-(z',\bz') \,,
\end{split}
\end{align}
where $G(z-z')$ is the Green's function for the transverse Laplacian given in \eqref{eq:Green_fcn}. Substituting this relation between $X^+$ and $P_-$ in \eqref{eq:hooft_final}, we obtain
\begin{align}
\begin{split}
	[X^+(z,\bz), X^-(z',\bz') ] &= 8\pi  i G_N G(z-z')\,.
\end{split}
\end{align}
This reflects an uncertainty in the $X^-, X^+$ coordinates of the shockwave and probe respectively in the quantum theory.

\section{Discussion}\label{sec:conclusion}

In this paper, we established a relation between the time delay acquired by a particle propagating in a shockwave background and the gravitational memory effect.  This implies that the memory mode, or equivalently the leading soft graviton mode, is related to the shockwave momentum introduced by 't Hooft in the context of scattering near a black hole horizon. Moreover, the canonically conjugate operator he postulated to correspond to the location of the black hole horizon may be identified with the Goldstone mode. Therefore, we are able to reinterpret the 't Hooft commutation relations as a simple consequence of the algebra of boundary operators derived from the covariant phase space formalism applied to asymptotically flat spacetimes. In particular, no black hole horizon needed to be present in our analysis.  This was foreseen in \cite{Verlinde:2019xfb,Verlinde:2022hhs} through the application of the 't Hooft commutation relations to a generic horizon of a causal diamond in flat space.

It is interesting to interpret this result from the perspective of soft gravitons. It is well-known that the memory effect can be thought of as a vacuum transition \cite{Strominger:2014pwa}, and this is associated with the production of soft gravitons. Here we see that this effect is directly related to the classical time delay of a particle crossing a shock, which is constrained by causality to be positive \cite{Shapiro:1964uw, Camanho:2014apa}.  By \eqref{eq:h-green}, \eqref{eq:time-delay}, and \eqref{eq:PN}, this should translate directly into a constraint on the memory effect (see \cite{Compere:2019rof} for a statement along these lines), and it would be interesting to further explore the implications of this relation. Moreover, the uncertainty in the location of the shockwave is tied to the Goldstone mode canonically conjugate to the soft graviton. Goldstone modes have previously been shown to appear in gravitational dressings \cite{Choi:2017bna, Arkani-Hamed:2020gyp, Nguyen:2021ydb, Kapec:2021eug, Kapec:2022hih}, suggesting a relation between shocks and coherent clouds of gravitons. However, this relation remains quite mysterious to us, and we hope to return to it in future work. 

Throughout this paper we restricted our analysis to the semiclassical regime: gravitational memory is a classical effect and we are only implementing canonical quantization. However, one can promote the stress tensor to a quantum operator.  This implies that the matter stress tensor can have non-zero {\it fluctuations} even in the vacuum where the {\it expectation value} vanishes.  That is, the two-point function of $T_{\mu\nu}^M$ is nonzero while the one-point vanishes, and this may correspond to nontrivial two-point functions involving the soft graviton and Goldstone modes $N$ and $C$, respectively \cite{Donnay:2020guq, Arkani-Hamed:2020gyp, Freidel:2022skz, Cotler:2023qwh}.  

Our results suggest that we may be able to probe quantum gravity by measuring the memory effect sourced by quantum fluctuations of spacetime rather than classical gravitational waves.  This is the essence of the proposal to detect observable effects in quantum gravity in \cite{Verlinde:2019xfb, Verlinde:2019ade, Zurek:2020ukz, Banks:2021jwj, Zurek:2022xzl, Gukov:2022oed, Verlinde:2022hhs,Zhang:2023mkf}.  
Of course, to have any observable effects from the quantum metric fluctuations, it is not enough for a single shockwave to be sourced by quantum fluctuations. Rather, the memory effect must {\it accumulate} sufficiently, over multiple quantum shocks, in order to be observable. Furthermore, to truly tie our results to an observational signature, we need to connect the time delay in the shockwave geometry and transverse displacement associated with the gravitational memory effect with a gauge invariant observable as measured by an interferometer. An idea for how this could happen was at the heart of the proposals in \cite{Verlinde:2019xfb,Zhang:2023mkf}. Nevertheless, by treating shockwaves as a gravitational memory effect, we open up additional directions to understand the infrared behavior of shockwave geometries, and we leave this study for future work.

\section*{Acknowledgements}

We thank Tim Adamo, Luis Apolo, Vincent Chen, Andrea Cristofoli, Laurent Freidel, Ben Freivogel, Vincent~S.~H. Lee, Prahar Mitra, Monica Pate, and Erik Verlinde for discussions, as well as Massimo Porrati and Andrew Strominger for useful feedback. We are supported by the Heising-Simons Foundation ``Observational Signatures of Quantum Gravity'' collaboration grant 2021-2817. The work of KZ is also supported by a Simons Investigator award and the U.S. Department of Energy, Office of Science, Office of High Energy Physics, under Award No. DE-SC0011632.

\appendix

\section{The 't Hooft Black Hole S-Matrix}
\label{app:S-matrix}

In this appendix, we review 't Hooft's argument for obtaining the phase shift \eqref{S-matrix} given the commutation relations \eqref{eq:commutators}. For notational simplicity we label all fields by just $z$ instead of $z,\bz$.

One can transition between eigenstates $|P^+ \rangle, |X_+\rangle$ of canonically conjugate variables $P^+$ and $X_+$
via the Fourier transform
\begin{equation}
\label{eq:Fourier}
	|\{X_+(z)\} \rangle = \CN \int \CD P^+ \exp{\left\{- i\int d^2 z' \gamma_{z'\bz'} P^+(z') X_+(z')\right\}} |\{P^+(z)\}\rangle\,,
\end{equation}
where $\{X_+(z)\}$ is shorthand for the formal product of $|X_+(z)\ra$ for all $z$ (and likewise for $\{P^+(z)\}$), $\CN$ a normalization constant, and we have normalized the states $|\{ P^+(z)\} \ra$ so that
\begin{equation}
\label{eq:normalization} 
	\langle \{P^+(z)\} | \{P'^+(z)\} \rangle = \CN' \prod_{z}  \delta\big(P^+(z) - P'^+(z)\big)\, ,
\end{equation}
where $\CN'$ is another normalization constant.
By \eqref{eq:gen-T} and \eqref{eq:h-green}, the  momentum operator is related to the stress tensor via 
\begin{equation} \label{eq:hooftP}
P_-(z) \equiv \int_{-\infty}^{x^-} dy^- \,T_{--}^{M(2)}(y^-, z) = -\frac{1}{16\pi G_N} \Box \amp(z) \,,
\end{equation}
for $x^{-} > x_0^-$.
Therefore \eqref{eq:Fourier} and \eqref{eq:normalization} imply that for some different normalization constant $\CN''$ 
\begin{equation}\label{eq:conj-relation}
	\langle \{X_+(z)\}| \{P^+(z)\} \rangle = \CN''\exp\bigg\{ i \int d^2 z'\,\gamma_{z'\bz'} P^+(z') X_+(z') \bigg\}\,.
\end{equation}
Finally, comparing with \eqref{S-matrix}, we see that we recover the phase shift \eqref{S-matrix} if we identify $|{\rm in}\ra \sim |\{P^+(z)\}\ra$, $|{\rm out}\ra \sim |\{X_+(z)\} \ra$, and
\begin{equation}\label{new-constr}
\begin{split}
	X_+(z) &= - 2\pi G_N \int d^2 z' \, \gamma_{z'\bz'}G(z - z') P^-(z') \\
	\implies \quad X^-(z) &= 4\pi G_N \int d^2 z' \, \gamma_{z'\bz'}G(z - z') P^-(z')\,,
\end{split}
\end{equation}
which corresponds precisely to \eqref{eq:xm-shift}. This is the time delay acquired by the source when propagating in the background of the probe. As such the commutation relations \eqref{eq:commutators} capture the quantum uncertainty in the positions of the source and probe introduced by their scattering \cite{tHooft:1996rdg}.

\section{Diffeomorphism Transforming the Shockwave Metric}
\label{app:diffeo}

In this appendix, we apply the diffeomorphism generated by $\xi$ from \eqref{eq:diffeo2} with $F$ given in \eqref{eq:F} to the shockwave metric \eqref{AS-met}, and show that we recover the metric \eqref{eq:partialbondi}. For convenience, we record here that the vector generating the diffeomorphism is
\begin{align}\label{eq:app-xi}
\begin{split}
	\xi &= F\,\p_u + \frac{1}{2}\Box F\,\p_r - \frac{1}{r}D^A F\,\p_A \,, \\
	(\Box + 2) F(u,z,\bz) &= \amp(z,\bz) \Th(u-u_0) - 2g(z,\bz) \,.
\end{split}
\end{align}
Using \eqref{eq:xi-indices}, we get
\begin{align}\label{eq:app-xi-lower}
	\xi_u = -\frac{1}{2}(\Box+2)F \,, \quad \xi_r =-F\,, \quad \xi_z = -r\p_z F\,.
\end{align}
Recall we chose $F$ to obey \eqref{eq:app-xi} such that we eliminate the $h_{uu}$ component in the shockwave metric \eqref{AS-met}. We now work out how the rest of the components transform. Recalling
\begin{align}
	\delta_\xi g_{\mu\nu} = \nabla_\mu\xi_\nu + \nabla_\nu\xi_\mu\,,
\end{align}
and that the nonzero Christofel symbols of the unperturbed metric are
\begin{align}
	\G^z_{rz} = \frac{1}{r}\,, \quad \G^z_{zz} = -\frac{2\bz}{1+z\bz}\,, \quad \G^u_{z\bz} = r\g_{z\bz}\,, \quad \G^r_{z\bz} = -r\g_{z\bz}\,,
\end{align}
we get 
\begin{align}\label{eq:app-gdiffeo}
\begin{split}
	\delta_\xi g_{ur} &= - \p_u F \,, \\
	\delta_\xi g_{uz} &= - r\p_u\p_z F - \frac{1}{2}\p_z (\Box+2)F \,, \\
	\delta_\xi g_{rz} &= 0 \,, \\
	\delta_\xi g_{zz} &= - 2 r D_z^2 F \,, \\
	\delta_\xi g_{z\bz} &= 0\,.
\end{split}
\end{align}
In particular, after performing the diffeomorphism, we have 
\begin{align}
\begin{split}
	& \TH{h}_{zz} \equiv r C_{zz} = \delta g_{zz} =  -2r D_z^2 F \quad\implies\quad C_{zz} = - 2 D_z^2 F \,.
\end{split}
\end{align}
Comparing with \eqref{eq:Czz-step}\,, which is 
\begin{align}\label{eq:app-Czz-soln}
\begin{split}
	C_{zz}(u,z,\bz) = D_z^2 N(z,\bz) \Th(u-u_0) - 2D_z^2 C(z,\bz)\,,
\end{split}
\end{align}
we can identify
\begin{align}\label{eq:app-F}
	F(u,z,\bz) = - \frac{1}{2}N(z,\bz)\Th(u-u_0) + C(z,\bz)\,.
\end{align}
Acting on both sides of this expression with $\Box+2$, we can derive from \eqref{eq:app-xi} that
\begin{align}\label{eq:app-hg}
	\amp(z,\bz) = - \frac{1}{2}(\Box+2) N(z,\bz) \,, \quad g(z,\bz) = - \frac{1}{2}(\Box+2) C(z,\bz)\,.
\end{align}
Finally, we observe that we can rewrite \eqref{eq:app-gdiffeo} in terms of $C$ and $N$ using \eqref{eq:app-F}, such that
\begin{align}\label{eq:app-metric-final}
\begin{split}
	\delta_\xi g_{ur} &= \frac{1}{2}N(z,\bz)\delta(u-u_0) \,, \\
	\delta_\xi g_{uz} &= \frac{r}{2}\p_z N(z,\bz)\delta(u-u_0) + \frac{1}{4}\p_z (\Box+2)N(z,\bz) \Th(u-u_0) - \frac{1}{2}\p_z(\Box+2)C(z,\bz) \\
	&= \frac{r}{2}\p_z N(z,\bz) \delta(u-u_0) + \frac{1}{2} D^z C_{zz}(u,z,\bz) \,, \\
	\delta_\xi g_{rz} &= 0 \,, \\
	\delta_\xi g_{zz} &= r D_z^2 \big( N(z,\bz)\Th(u-u_0) - 2C(z,\bz) \big)  \\
	&= rC_{zz}(u,z,\bz) \,, \\
	\delta_\xi g_{z\bz} &= 0\,.
\end{split}
\end{align}
In terms of these fields, the resulting metric is 
\begin{align}\label{post-diffeo}
\begin{split}
	ds^2 &= -du^2 - 2 \,du\,dr + 2r^2\gamma_{z\bz}\,dz\,d\bz   \\
	& \qquad + \big( r C_{zz}(u,z,\bz)\,dz^2 + \cc \big) +  \big(  D^zC_{zz}(u,z,\bz)  \,du\,dz + \cc \big)  \\
	&\qquad + N(z,\bz)\delta(u-u_0)\,du\,dr +  \big( r\p_z N(z,\bz)\delta(u-u_0)\,du\,dz + \cc \big) \,,
\end{split}
\end{align}
which is precisely \eqref{eq:partialbondi}.

Finally, we briefly comment on another metric that is diffeomorphic to the shockwave metric \eqref{AS-met}. Consider a diffeomorphism generated by the vector
\begin{align}
	\chi =  \frac{\amp(z,\bz)}{2}\Th(u-u_0)\p_r \,.
\end{align}
Using \eqref{eq:xi-indices}, this means
\begin{align}
	\chi_u = -\frac{\amp(z,\bz)}{2}\Th(u-u_0)\,, \qquad \chi_r = 0 \,,\qquad \chi_z = 0\,.
\end{align}
Performing the diffeomorphism on \eqref{AS-met}, we obtain the metric
\begin{align}
\begin{split}
	ds^2 &= - du^2 - 2du\,dr + 2r^2\g_{z\bz}\bigg( 1 - \frac{2}{r} \chi_u(u,z,\bz)  \bigg)dz\,d\bz \\
	&\qquad  + 2\big(\p_z \chi_u(u,z,\bz) \,du\,dz + \cc \big) \,.
\end{split}
\end{align}
In the limit where $r \to \infty$, this reduces precisely to the metric studied in \cite{Verlinde:2022hhs}.

\section{Symplectic Form of Transformed Metric}\label{app:symplectic}

In this appendix, we compute the symplectic form of the transformed shockwave metric given in \eqref{post-diffeo}, namely
\begin{align}\label{symp_metric0}
\begin{split}
	ds^2 &= -du^2 - 2 \,du\,dr + 2r^2\gamma_{z\bz}\,dz\,d\bz   \\
	& \qquad + \big( r C_{zz}(u,z,\bz)\,dz^2 + \cc \big) +  \big(  D^zC_{zz}(u,z,\bz) \,du\,dz + \cc \big)  \\
	&\qquad + \sigma(u,z,\bz)\,du\,dr +  \big( r\p_z \sigma(u,z,\bz)\,du\,dz + \cc \big)\,,
\end{split}
\end{align}
where
\begin{align}
\label{eq:sigma}
	\sigma(u,z,\bz) \equiv N(z,\bz)\delta(u-u_0)\,.
\end{align}
Following \cite{Barnich:2007bf, Compere:2016jwb, Compere:2018ylh, Alessio:2019cae}, the symplectic form for a metric in linearized gravity is given by (we denote $h \equiv h_\mu{}^\mu$)
\begin{align}
\label{eq:symplectic-form}
\begin{split}
	\Omega_\Sigma[\delta_1 h,\delta_2 h] &= \frac{1}{16\pi G_N} \int_{\Sigma} \sqrt{-g} \w^\mu \,d\Sigma_\mu \,, \\
	\w^\mu &= \frac{1}{2} \delta_2 h \nabla^\mu \delta_1 h + \delta_2 h_{\nu\rho} \nabla^\nu \delta_1 h^{\mu\rho} - \frac{1}{2} \delta_2 h \nabla_\nu \delta_1 h^{\mu\nu} - \frac{1}{2} \delta_2 h^{\nu\rho}\nabla^\mu \delta_1 h_{\nu\rho} \\
	&\qquad - \frac{1}{2}\delta_2 h^{\mu\rho} \nabla_\rho \delta_1 h - (1 \leftrightarrow 2) \,, 
\end{split}
\end{align}
where $\w^\mu$ is the symplectic current, $\Sigma$ a Cauchy slice, $ d\Sigma_\mu$ the volume 3-form labeled by its orthogonal vector $x^\mu$, and $\delta h^{\mu\nu} \equiv g^{\mu\sigma} g^{\nu\rho} \delta h_{\sigma \rho}$. For the case we are interested in, we want to take the Cauchy slice $\Sigma$ to be $\ci^+$, which is defined as the $u+2r = v_0$ hyperplane with $v_0 \to \infty$. This means that $n_{\mu} = (\frac{1}{2}, 1, 0, 0)$, or equivalently that the normal vector is $n^\mu = (-1, \frac{1}{2},0,0)$. The relevant components of the symplectic current are therefore $\omega^u$ and $\omega^r$. It is straightforward to show that $\omega^u = \mathcal{O}(r^{-3})$ and hence its contribution to the symplectic form vanishes in the limit $r \rightarrow \infty$. We are left with evaluating $\omega^r$. 

To this end, we start by computing the nonzero components of the linearized metric 
\begin{align}
	h_{ur} = \frac{1}{2} \sigma \,, \quad h_{zz} = r C_{zz}\,, \quad h_{uz} = \frac{r}{2} D_z \sigma + \frac{1}{2}D^z C_{zz} \,.
\end{align}
Since $\sqrt{-g} = r^2\g_{z\bz}$, we will be interested in the terms of $\w^r$ that fall off slower than $r^{-3}$. Keeping only terms quadratic in $h$, we first focus on the terms in $\w^r$ that could lead to divergences when multiplying $\sqrt{-g}$. We find
\begin{align}
\begin{split}
	\lim_{r\to\infty} \sqrt{-g} \w^r \Big|_{\text{div terms}} &= - \gamma_{z\bz} \bigg[ \frac{r}{4}D_A \delta_2 \sigma D^A \delta_1 \sigma  + \frac{r}{4}\delta_2 \sigma D_A D^A \delta_1 \sigma +  r \delta_2 \sigma \delta_1 \sigma   \bigg] -(1 \leftrightarrow 2) \,.
\end{split}
\end{align}
Upon integration by parts over the $z,\bz$ coordinates and antisymmetrization, this term drops out.

We conclude that the divergent terms cancel out and the symplectic form in the large $r$ limit reduces to
\begin{align}
\begin{split}
	\lim_{r\to\infty} \sqrt{-g} \w^r &= -\g_{z\bz} \bigg[ \frac{1}{2}  D_A \delta_2 \sigma D_B \delta_1  C^{BA} + \frac{1}{2} \delta_2 C^{BA} D_B D_A \delta_1 \sigma + \frac{1}{4} \delta_2 \sigma D_A D_B \delta_1 C^{AB} \\
	&\qquad\qquad + \frac{1}{4} D_A  \delta_2 C^{AB} D_B \delta_1 \sigma - \frac{1}{2} \delta_2 C^{AB} \p_u \delta_1 C_{AB} \bigg] - (1  \leftrightarrow 2) \,.
\end{split}
\end{align}
Note that after integration by parts and dropping total derivatives on the sphere, the first four terms simplify and we are left with
\begin{align}
\begin{split}
	\lim_{r\to\infty} \sqrt{-g}\w^r  &= -\frac{1}{2} \gamma^{z\bz} \big( \delta_1 C_{zz} \delta_2 N_{\bz\bz} + \delta_1 C_{\bz\bz} \delta_2 N_{zz} \big) \\
	&\qquad + \frac{1}{2}\gamma^{z\bz}\left(\delta_1 C_{zz} D_{\bz}^2 \delta_2 \sigma + \delta_1 C_{\bz\bz} D_z^2 \delta_2\sigma \right) - (1 \leftrightarrow 2)\,.
\end{split}
\end{align}
Defining 
\begin{equation}
\hat{N}_{AB} \equiv N_{AB} - D_A D_B \sigma \,,
\end{equation}
the symplectic form becomes
\begin{align}\label{symp_final}
\begin{split}
	\Omega_{\ci^+}[\delta_1 h, \delta_2 h] &= \frac{1}{16\pi G_N} \int_{\ci^+} \sqrt{-g} \w^r n_r \,du\,dz\,d\bz \\
	&= -\frac{1}{32\pi G_N} \int_{\ci^+} \Big[ \g^{z\bz} \big( \delta_1 C_{zz} \delta_2 \hat{N}_{\bz\bz} + \delta_1 C_{\bz\bz} \delta_2 \hat{N}_{zz}  \\
	&\qquad - \delta_1  \hat{N}_{\bz\bz} \delta_2 C_{zz} - \delta_1 \hat{N}_{zz} \delta_2 C_{\bz\bz} \big) \Big]du\,dz\,d\bz \\
	&= -\frac{1}{16\pi G_N} \int_{\ci^+} \Big[ \g^{z\bz} \big( \delta_1 C_{zz} \delta_2  \hat{N}_{\bz\bz} - \delta_2 C_{zz} \delta_1 \hat{N}_{\bz\bz}    \big) \Big]du\,dz\,d\bz + \cdots \,,
\end{split}
\end{align}
where we integrated by parts to obtain the last equality, and $\cdots$ denote additional boundary terms. From this we can immediately read off the Poisson bracket involving the bulk degrees of freedom to be 
\begin{align}
	\big\{ C_{\bz\bz}(u,z,\bz), \hat{N}_{ww}(u',w,\bw) \big\} = 16 \pi G_N \gamma_{z\bz} \delta(u-u') \delta^{(2)}(z-w)\,.
\end{align} 
For $\sigma = 0$, this agrees with  \eqref{eq:Poisson-CN} as expected. For $\sigma$ given in \eqref{eq:sigma}, which was obtained from the shockwave metric \eqref{AS-met} via the diffeomorphism constructed in Section~\ref{ssec:diffeo} and $N_{AB}$ entirely ``soft'', i.e. associated with the shear in \eqref{eq:Czz-step}, the symplectic form \eqref{symp_final} vanishes. This is consistent with the vanishing of the symplectic form for shockwave metrics of the form \eqref{AS-met}, which can be straightforwardly verified.

\normalem

\bibliography{VZeffect-bib}{}
\bibliographystyle{utphys}

\end{document}